\documentclass[12pt]{article}

\usepackage{a4}
\usepackage{titlesec}
\usepackage{harvard}

\usepackage{endnotes}

\usepackage{amsthm}

\theoremstyle{definition}

\usepackage[dvipsnames]{xcolor}

\usepackage{graphicx}

\usepackage{url}

\usepackage{amsmath}
\usepackage{amssymb}

\newcommand{\dd}{\mathrm{d} }
\newcommand{\St}{\mathrm{I}}

\titleformat{\section}
  {\normalfont\Large\bfseries}{\thesection}{1em}{}

\titleformat{\subsection}
  {\normalfont\large\itshape}{\thesubsection}{1em}{}

\title{\textbf{\Large{The Concept of Average Loop Impact as a System-Wide Measure of Feedback Dominance Analysis}}}

\normalsize

\author{\textbf{John Hayward}  \\\\\\\
School of  Computing and Mathematics\\ University of South Wales\\
Pontypridd, CF37 1DL, Wales, UK \\\\
john.hayward@southwales.ac.uk}

\citationstyle{dcu}

\begin{document}

\maketitle

\begin{abstract}
In system dynamics, the concept of loop impact measures the effect of a feedback loop on the curvature in the behaviour of the stocks in the loop \cite{hayward2014model}. It is a ratio measure of the acceleration imparted by the loop into a stock and can also be thought of as a ratio measure of the force exerted between adjacent stocks in the loop \cite{hayward2017newton}.  As such, a loop of $n$ stocks is described by $n$ impacts. Loop impact is used to examine regions of loop dominance in stock behaviour. Sometimes, it is helpful to present a dominance analysis for a complete system rather than an individual stock. This paper introduces the concept of \emph{average loop impact} to explore such a system-wide  loop dominance analysis.
\end{abstract}

\section*{Introduction}
In recent years, the system dynamics field has seen a growing number of methods and tools for loop dominance analysis. The methods group into two broad categories, eigenvalue analysis and pathway approaches. See \citeasnoun{duggan2013special} and \citeasnoun{kampmann2020analytical} for a review of the field. 

In 2020, a new tool became available in the Stella Architect software platform of isee systems inc, using the Loops that Matter method \cite{schoenberg2020understanding}. Loops that Matter is a pathway method, with some similarities to the Pathway Participation Metric (PPM) \cite{mojtahedzadeh2004using} and Loop Impact \cite{hayward2014model}, but differs from these by defining a single measure for a feedback loop with two or more stocks, enabling a system-wide analysis of dominance. By contrast, the closely related methods of PPM and Loop Impact analyse the behaviour of loops on individual stocks only and as yet do not have a single measure for higher-order loops.  In this technical note, a single measure of a feedback loop called \emph{Average Loop Impact} will be introduced and compared with the Loop Impact method on individual stocks using three simple SD models. 

\begin{figure}[!ht]
          \begin{center}
   \includegraphics[width=9cm] {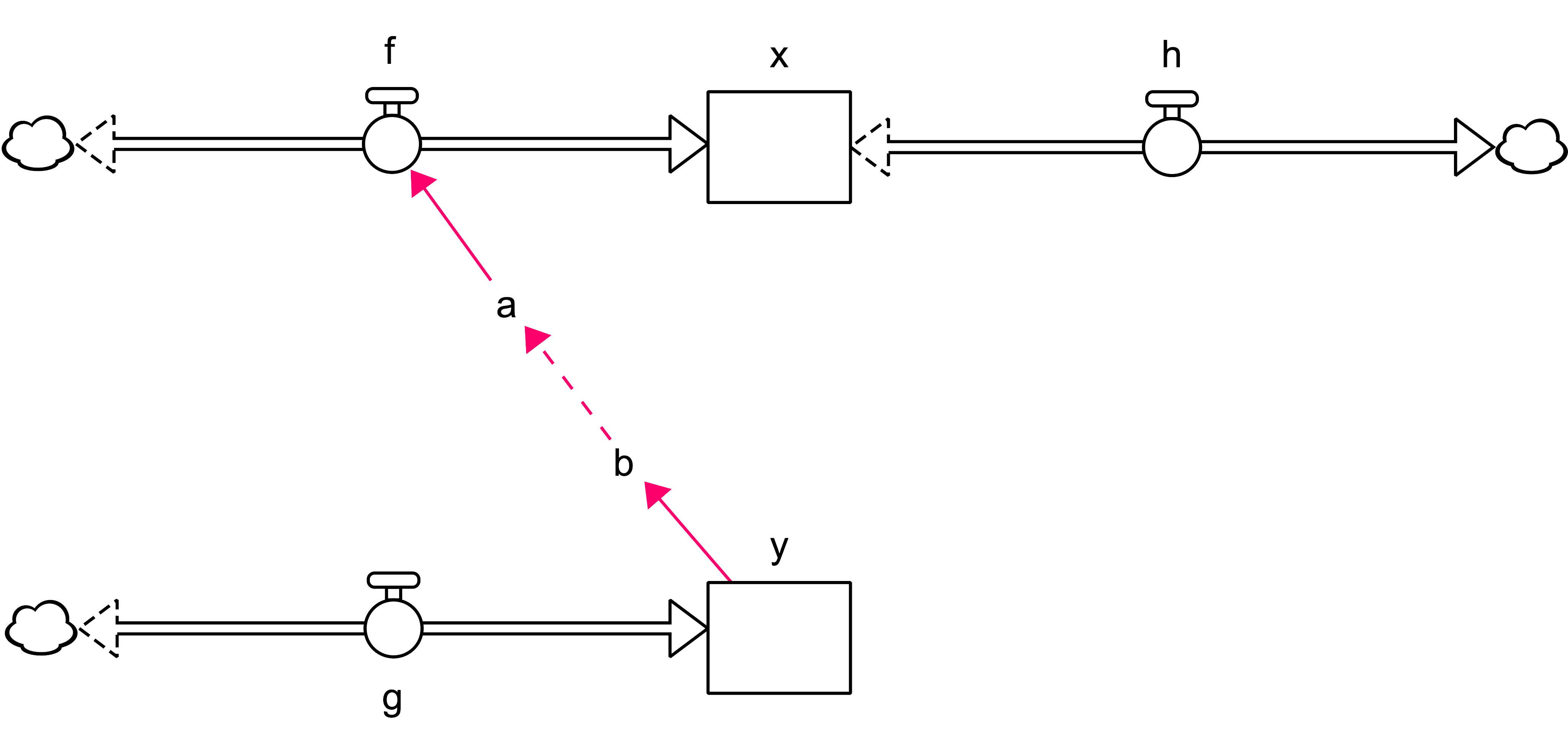}
       \end{center}
    \vspace{-20 pt}
    \caption{\small{One stock $x$ influenced by another stock $y$.}} \label{figure1.fig}
 \end{figure}

Loop dominance analysis connects model structure to behaviour. For the pathway methods, the key structures are the causal links between stocks, flows and auxiliaries, which can be defined analytically and computed numerically. These definitions will determine the behaviour to be explained Consider two adjacent stocks with intermediate auxiliaries, figure \ref{figure1.fig}.

 PPM and loop impact analyse behaviour by examining the acceleration of a target stock $x$ due to changes in source stock $y$ using the second derivative of $x$ \cite[eq.~5]{mojtahedzadeh2004using}, \cite[eq.~2]{hayward2014model},  \cite[eq.A4]{hayward2019force}. For the causal chain in figure \ref{figure1.fig}, where $\dot{x}=f(y)$
 \begin{equation}
 \frac{\dd^2x}{\dd t^2}=\left(\frac{\partial f}{\partial y} \frac{\dot{y}}{\dot{x}}\right) \dot{x}  + \dot{h}\label{one.eq}
 \end{equation} 
 The \emph{impact} of $y$ on $x$ is the coefficient of $\dot{x}$ in (\ref{one.eq}) :
 \begin{equation}
 \St_{\underline{yx}}=\frac{\partial f}{\partial y} \frac{\dot{y}}{\dot{x}} \label{two.eq}
 \end{equation}
 If $f$ had been an outflow, the formula would have a minus sign. Impact measures the degree of curvature in the behaviour of a stock in units of \emph{per unit time}, independent of the stock units \cite{hayward2014model}. If there are many links to the stock, each of which is part of a feedback loop, PPM sums up all the raw impacts and define the pathway participation metric for one connection as its proportion of the total, \cite[eq.~6]{mojtahedzadeh2004using}. The loop impact is the raw equivalent of PPM.
 
\begin{figure}[!ht] 
          \begin{center}
   \includegraphics[width=6.5cm] {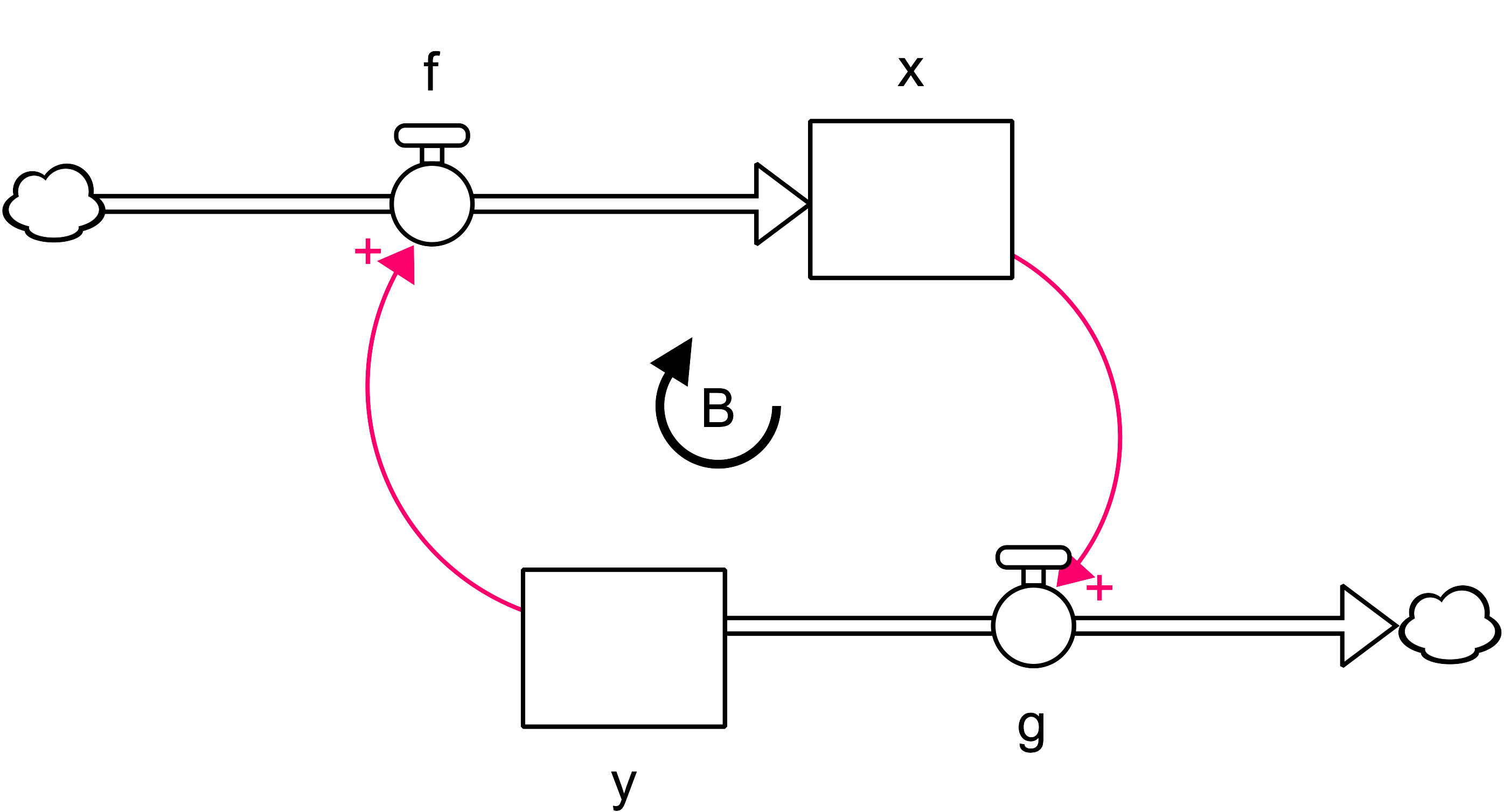}
       \end{center}
    \vspace{-20 pt}
    \caption{\small{Single second-order balancing loop.}} \label{figure2.fig}
 \end{figure}

 If there are several stocks in a loop, then there is a loop impact or PPM for each stock. For example, for the second-order loop in figure \ref{figure2.fig}, loop $B$ has two impacts, one each for $x$ and $y$. Likewise, $B$ would have two PPMs. However,  there would need to be other loops through these stocks for the PPMs to be less than 100\%. 
 
The definition of impact (\ref{two.eq}) has two parts. Firstly, there is the partial derivative $\partial f/\partial y$ that represents the chain of \emph{same way} and \emph{opposite way} links from $y$ to the flow of $x$ given by a product of partial derivatives of all the elements -- the link gain \cite{richardson1995loop,kampmann2012feedback}.  Secondly, there is the ratio of the rates of change of the stocks $\dot{y}/\dot{x}$ that weights the link gain with the relative changes in the stocks. In first-order loops, this weighting is unity as the target stock is its own source. Thus, the loop impact is the loop gain. In loops with two or more stocks, the weighting apportions the loop gain between the inter-stock links in the loop such that the product of the impacts is the loop gain \cite[appendix~C]{hayward2014model}.

Unlike the Loops that Matter method, the Loop Impact method and PPM do not provide a single measure for a loop of two or more stocks. The Loops that Matter method is able to obtain a single score for such loops by multiplying all the links together \cite{schoenberg2020understanding}. This procedure is not helpful using loop impact as such a product merely returns the loop gain, which on its own does not capture the effects of loops on stocks or other elements in the loop. Instead, the  \emph{average loop impact} is introduced, whose magnitude is the average of the absolute values of all impacts in the loop, and whose sign matches that of the loop gain.  For example, for the loop in figure \ref{figure2.fig}, the average loop impact is $-(|\St_{\underline{yx}}| + |\St_{\underline{xy}}|)/2$. Although this average does not correspond to any single graphical behaviour over time, it does measure, in units \emph{per unit time}, the amount of curvature induced in all stocks influenced by that loop in such a way that loops with many stocks are not over-weighted. If the influence on one stock was slight, the weighting for the loop would be reduced.  This definition easily extends to PPM.

Next, three models are analysed using individual loop impacts and the average.

\section*{Limits-to-Growth Model}
The first model is a  one stock limits-to-growth model  with an added draining process, preventing the stock from achieving the natural carrying capacity. The model has one reinforcing loop driving the growth and two balancing loops in opposition.  Loop impacts are computed numerically using \possessivecite{hayward2014model} method and confirmed  analytically using pathway differentiation (\citename{hayward2017newton}, \citeyear*{hayward2017newton,hayward2019force}).  There is no need to consider average loop impact here as there is only one stock in each loop. This model is used to demonstrate how the loop impact method works where there are no higher-order loops. The results are displayed in figure \ref{figure4.fig} for two different rates of business demolition.

 \begin{figure}[!ht]
          \begin{center}
   \includegraphics[width=12cm] {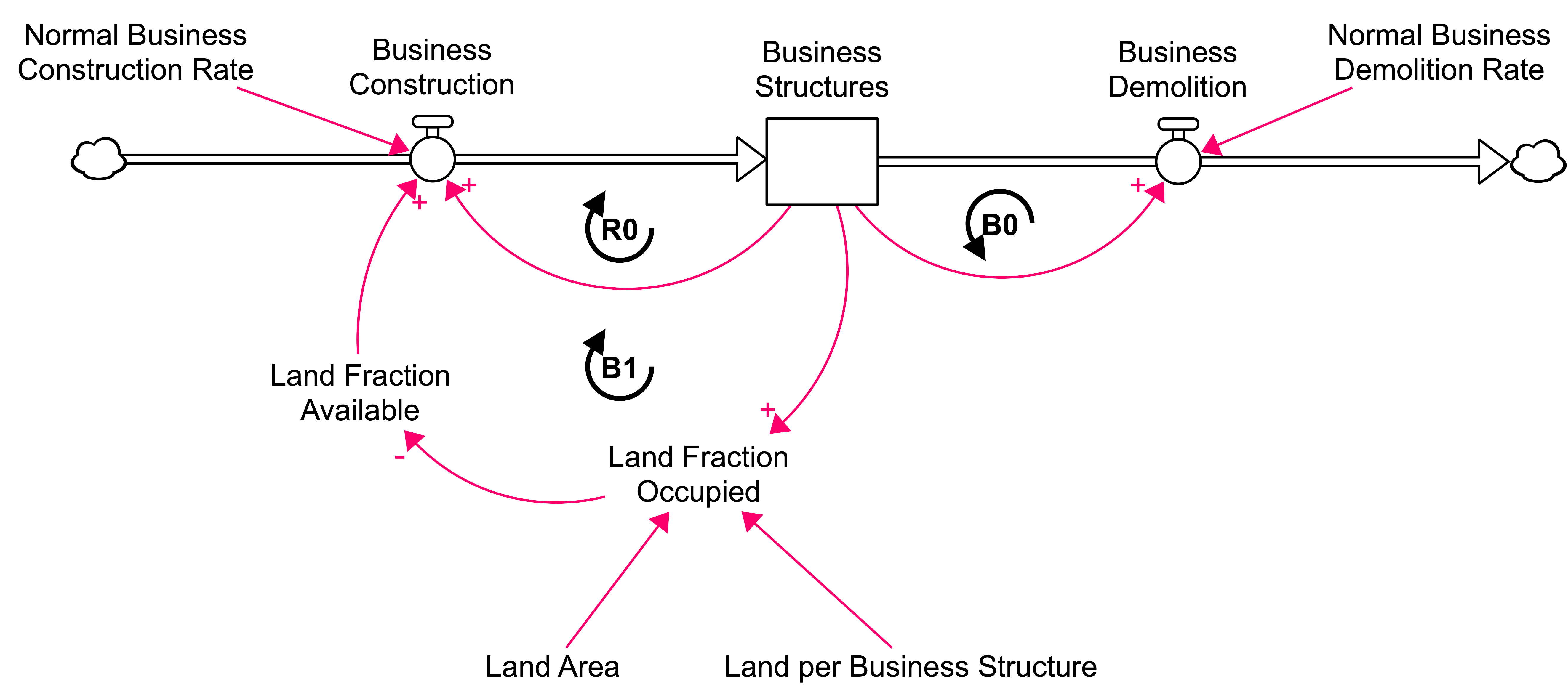}
       \end{center}
    \vspace{-20 pt}
    \caption{\small{Three-loop Limits-to-Growth model.}} \label{figure3.fig}
 \end{figure}

 \begin{figure}[!ht]
          \begin{center}
   \includegraphics[width=9cm] {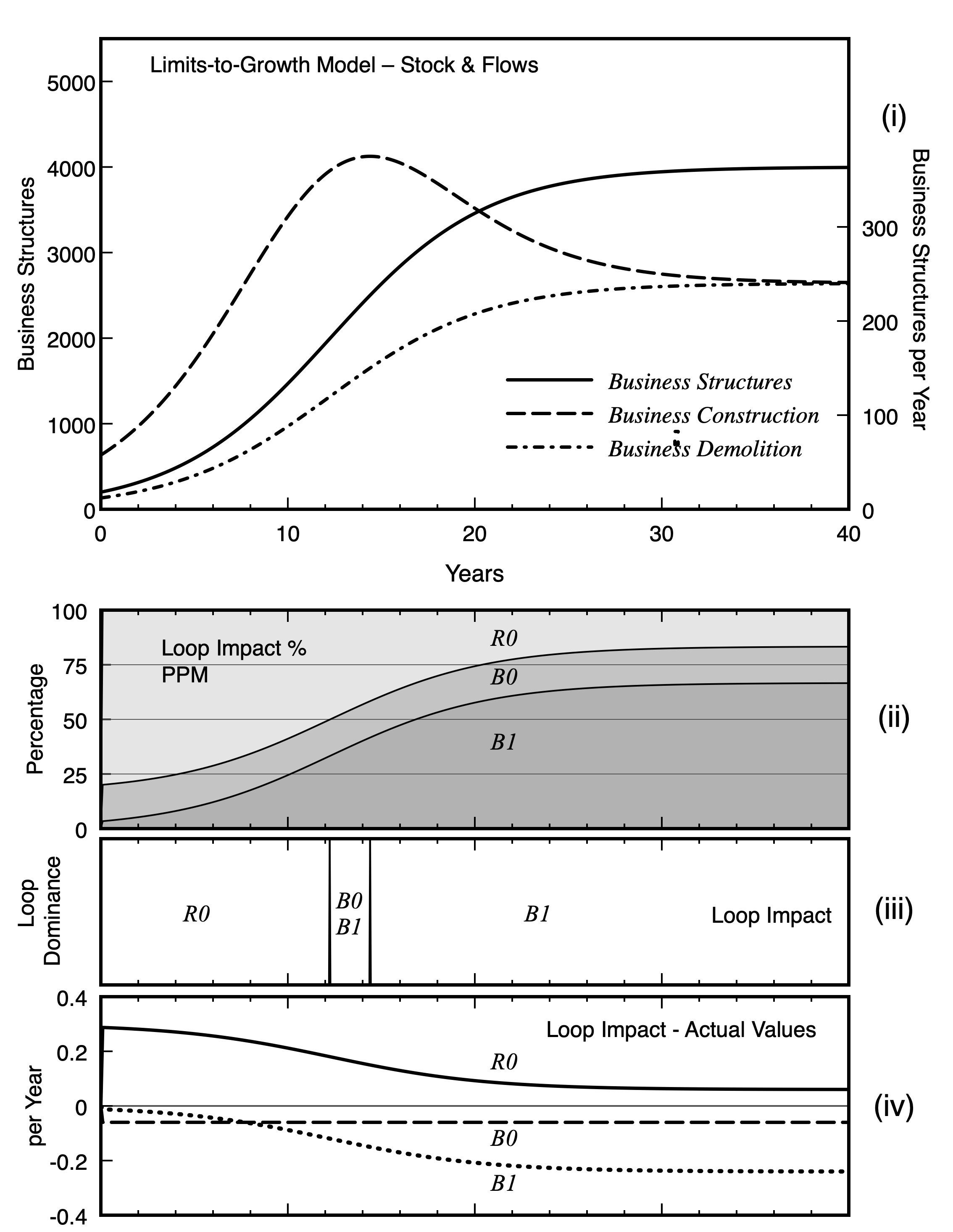}
       \end{center}
    \vspace{-20 pt}
   \caption{\small{ Results of the three-loop Limits-to-Growth model. (i) Stocks and flows; (ii) Loop impacts as a percentage of influence on the stock (PPM); (iii) Loop dominance transitions for loop impact; (iv) Loop impact, actual values. \emph{business construction rate }=0.3, \emph{business demolition rate} = 0.06, \emph{land area} = 1000, \emph{land per business structure} = 0.2, initial \emph{Business Structures} = 200.  }}  \label{figure4.fig}
 \end{figure}

For a demolition rate of 0.06, the stock value falls well short of capacity, figure \ref{figure4.fig}(i). The inflow business construction rises to a maximum value, then declines until the inflow and outflow match. The loop impact approach explains the acceleration phase of stock growth by the dominance of $R_0$, with the deceleration phase, after the inflexion point, explained the balancing loops being dominant,  figure \ref{figure4.fig}(iii). There is a brief phase where it takes both balancing loops to slow the growth, but for these parameter values, the final approach to equilibrium is due to the capacity loop $B_2$ alone. The percentage values of the loop impact, \ref{figure4.fig}(ii), shows that the influence of $B_0$ remains constant throughout. These percentages are the Pathway Participation Metric of \citeasnoun{mojtahedzadeh2004using}. 

The reason why loop $B_0$ has a constant proportional influence on the stock is because it is a linear loop and has constant gain. Thus, it has a constant impact, which is equal to the demolition rate \citeaffixed{hayward2017newton}{compare with eq.~11 in}. The actual values of the impacts are given in figure \ref{figure4.fig}(iv). Initially the growth loop $R_0$ has the highest value of impact, but it declines throughout as the stock gets closer to capacity. By contrast, the capacity loop $B_1$ increases in value throughout. Initially, it is numerically smaller than the demolition loop $B_0$, but for most of the run is larger. Although $B_0$ is only used in a brief period in the dominance analysis, figure \ref{figure4.fig}(iii), its importance is that it lowers the equilibrium value of the stock. The smaller its gain, the smaller its influence on the dominance analysis, and the closer the final value of the stock to the carrying capacity.

It is noted that the loop impacts as computed here describe the curvature of stock behaviour only. They do not directly describe the behaviour of the two flows. It could be argued that $R_0$ causes \emph{Business Construction} to rise, and dominance by $B_1$ causes it to fall, figure \ref{figure4.fig}(i). But they do not explain the \emph{curvature} in flow behaviour. To explain that, separate impacts of the loops on the flows need to be computed from considering the third derivative of the stocks, see for example,  \citeasnoun{hayward2017loop}.

\section*{Inventory-Workforce Model}
\begin{figure}[!ht]
          \begin{center}
   \includegraphics[width=12cm] {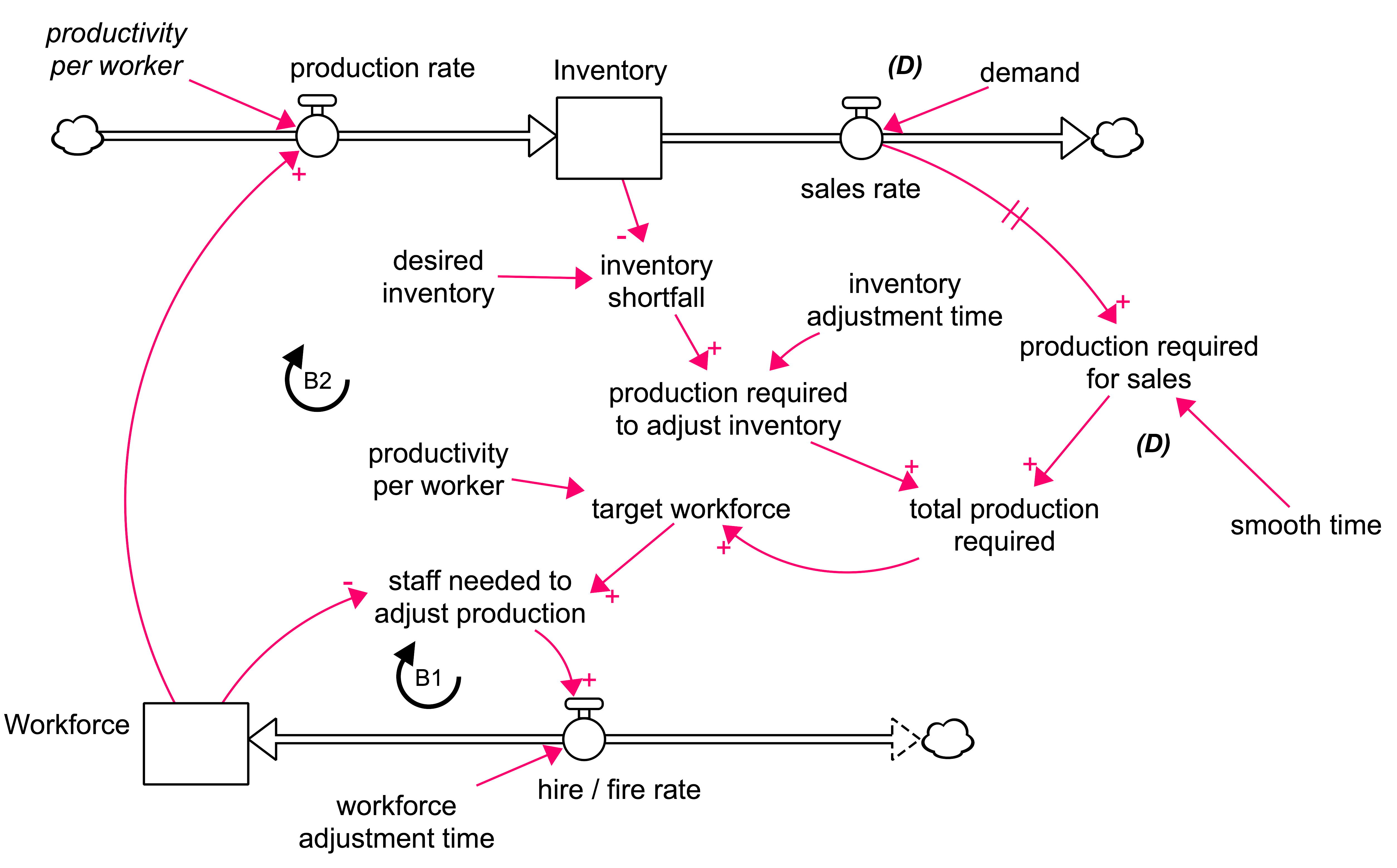}
       \end{center}
    \vspace{-20 pt}
    \caption{\small{Inventory--Workforce model.}} \label{figure5.fig}
 \end{figure}

\begin{figure}[!ht]
 \begin{minipage}[b]{0.5\linewidth}
\centering
   \includegraphics[width=9cm] {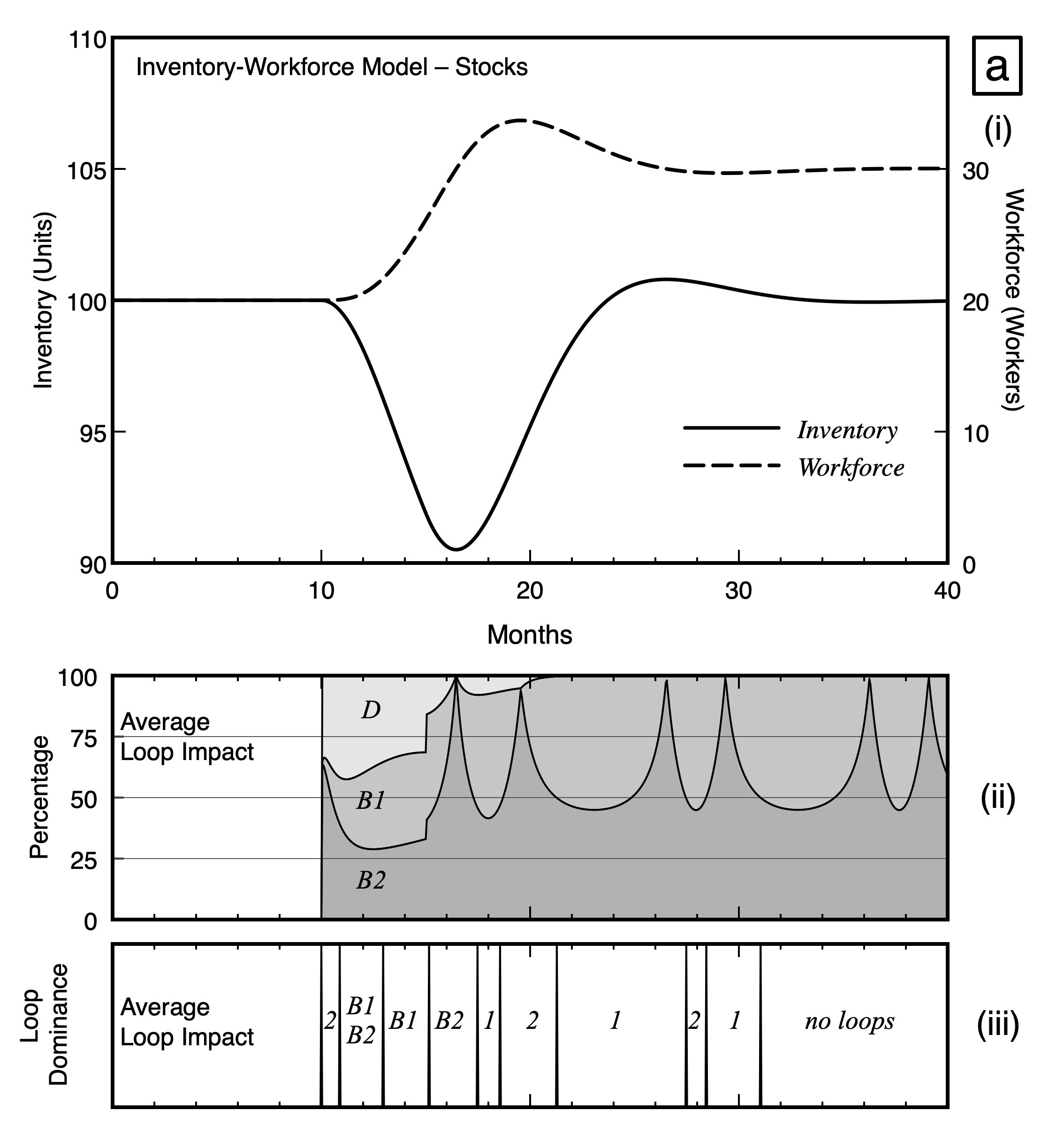}
      \end{minipage}
\hspace{0.1cm}
\begin{minipage}[b]{0.5\linewidth}
\centering
   \includegraphics[width=9cm] {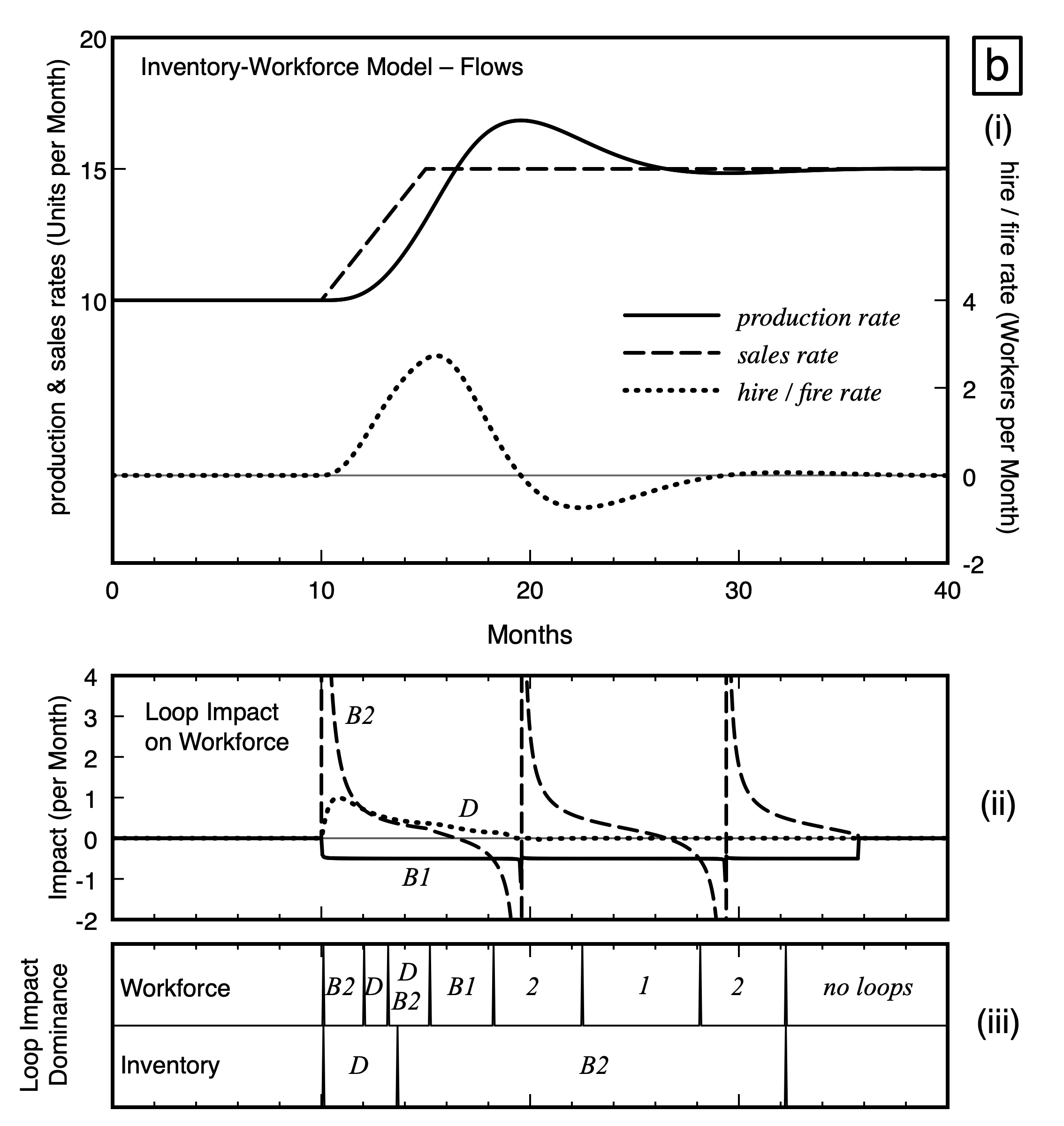}
      \end{minipage}
      \vspace{-25 pt}
       \caption{\small{Results of the Inventory-Workforce model.  (a) (i) Stocks; (ii) Loop impacts as a percentage of influence on the stock; (iii) Loop dominance transitions for loop impact. (b). (i) Flows; (ii) loop impacts on \emph{Workforce}; (iii) Loop dominance transitions for separate loop impact on the stocks;   \emph{productivity per worker} = 0.5, \emph{inventory adjustment time} = 3, \emph{workforce adjustment time} = 2, \emph{smooth time} = 2, \emph{initial workforce} = 20, \emph{initial inventory} = 100.}} \label{figure6.fig}
   \end{figure}

Our second example is a two-stock inventory-workforce model figure \ref{figure5.fig}. The model has two stocks, and two feedback loops –– one first-order $B_1$ and the other second-order $B_2$. Additionally, there is an exogenous demand with a smoothed delay between demand and the total production required. This model has been previously analysed using loop impact \cite{hayward2017newton}. The loop impact method is extended using average loop impact for the two stocks in $B_2$ and compare it with impacts on the individual stocks. Demand influences both stocks. Their impacts can also be averaged to provide a single measure of exogenous influence. The results are given in figure \ref{figure6.fig}.

 Exogenous demand changes at month 10, increasing from 10 units a month to 15 units a month in 3 months, after which it remains at the higher value,  figure \ref{figure4.fig}b(i). From month 10, the system undergoes damped oscillations, with the inventory stabilising on its desired value and the workforce reaching a higher equilibrium, figure \ref{figure6.fig}a(i). The percentage influences using average loop impact also oscillates, figure \ref{figure6.fig}a(ii). The dominance changes between $B_1$ and $B_2$ in average loop Impact figure\ref{figure6.fig} a(iii), similar to the changes in impact dominance on the workforce, figure\ref{figure6.fig} b(iii).  However, changing the adjustment times will produce scenarios where $B_2$ dominates throughout in average loop impact. 
 
 The percentage influences in Loop Impact exhibits sharp corners, figure \ref{figure6.fig}a(ii). This behaviour is an artefact of computing proportions where one of the raw measures is briefly infinite. Infinities in loop impact occur at turning points in the stock values. There are three such points for the impact of $B_2$ on \emph{Workforce}, figure \ref{figure6.fig}b(ii). Likewise, there are three such infinities in the impact of $B_2$ on \emph{Inventory}, which occur when its impact on \emph{Workforce} is zero. Because average loop impact uses impact on both stocks, its  infinities occur in pairs, figure \ref{figure6.fig}a(ii).

The advantage of average loop impact is that it gives a single number to measure the effect of a loop, enabling a single dominance analysis for a whole connected system figure \ref{figure6.fig}a(iii).  However, the disadvantage is that the connection between structure and behaviour is complex; multiple stock behaviour has to be inferred from a single dominance transition diagram. By contrast, using loop impacts for each stock separately has a clear interpretation in terms of the curvature of stock behaviour. For example, for the inventory-workforce model, the acceleration and deceleration in workforce numbers  is explained by its initial response to demand $D$ directly through the increase production required for sales and the changing inventory shortfall via $B_2$, figure \ref{figure6.fig}b(iii), \citeaffixed{hayward2017newton}{see also}. Subsequent oscillations are determined by transitions between the restoring feedback of $B_2$ and the frictional force of $B_1$. Apart from the initial direct effect of demand, the inventory's behaviour is explained solely by $B_2$ -- a reaction to the behaviour of the workforce. Loops with more than one stock require more than one representative number to  keep the link between behaviour and structure clear. Thus, the system is easier to explain using two loop impacts and the Newtonian Interpretive Framework of \citeasnoun{hayward2017newton}, rather than the single number of average loop impact.

\section*{SIR Model}
\begin{figure}[!ht]
          \begin{center}
   \includegraphics[width=11cm] {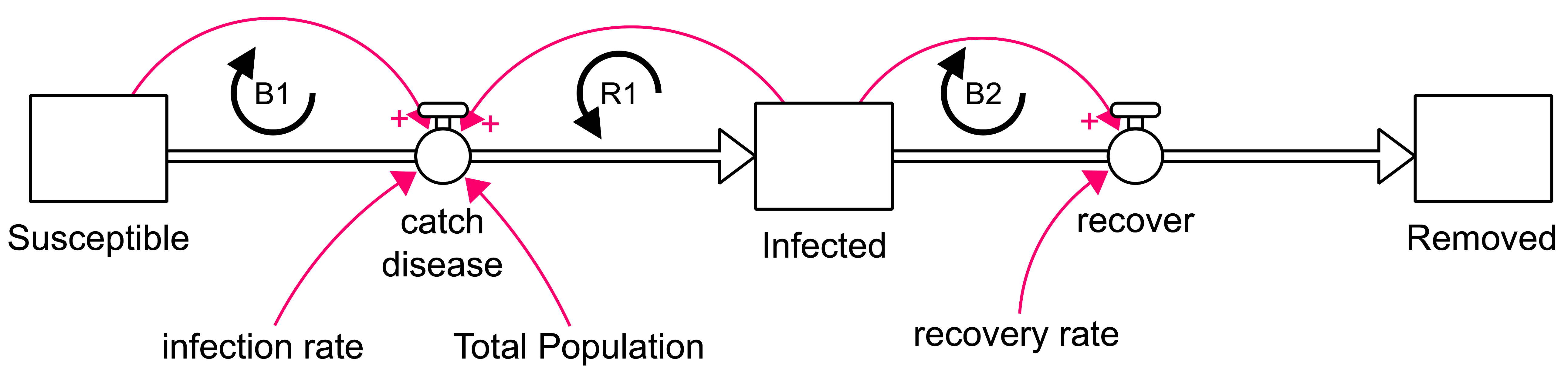}
       \end{center}
    \vspace{-20 pt}
    \caption{\small{SIR model.}} \label{figure7.fig}
 \end{figure}

\begin{figure}[!ht]
 \begin{minipage}[b]{0.5\linewidth}
\centering
   \includegraphics[width=9cm] {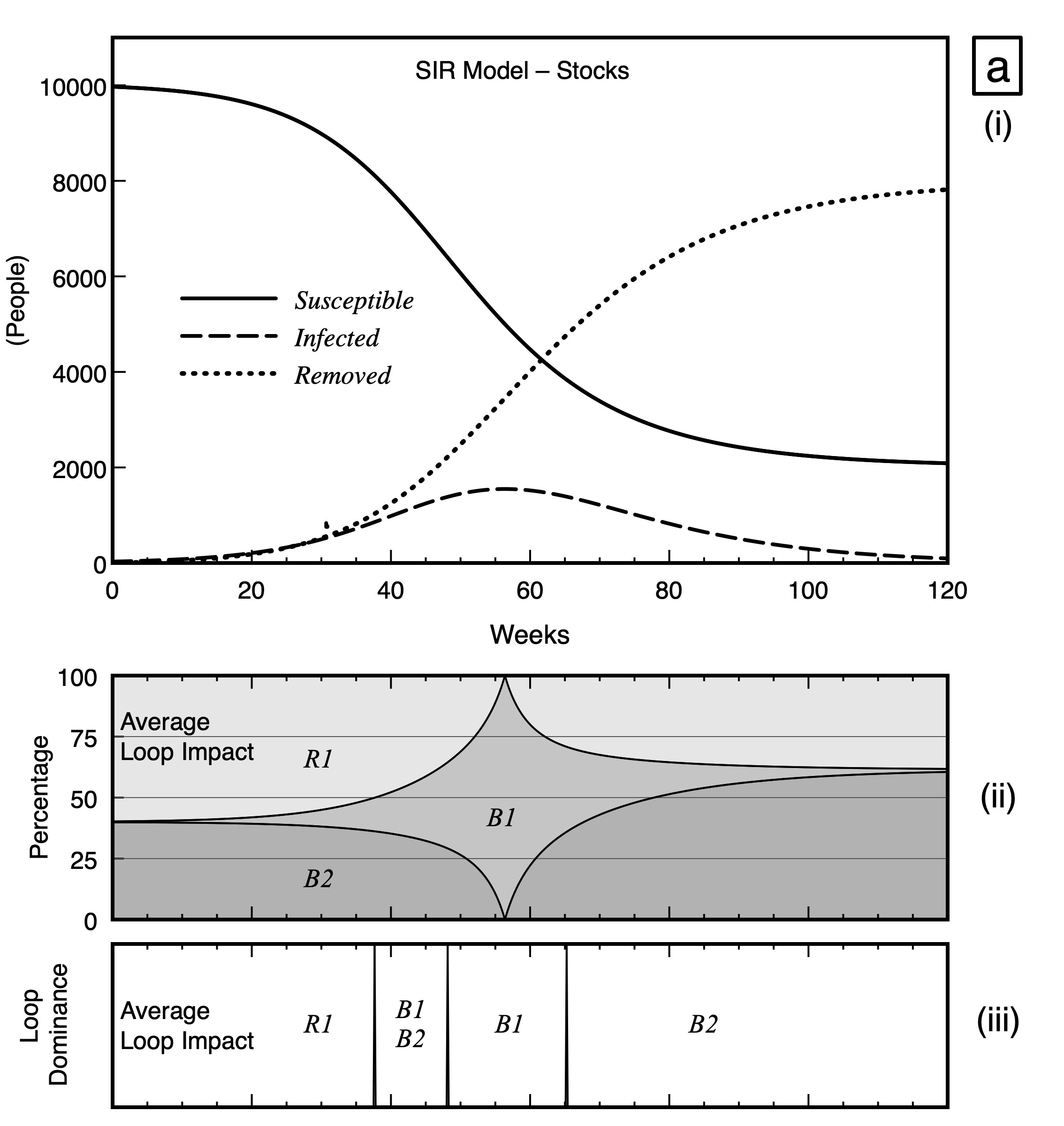}
      \end{minipage}
\hspace{0.1cm}
\begin{minipage}[b]{0.5\linewidth}
\centering
   \includegraphics[width=9cm] {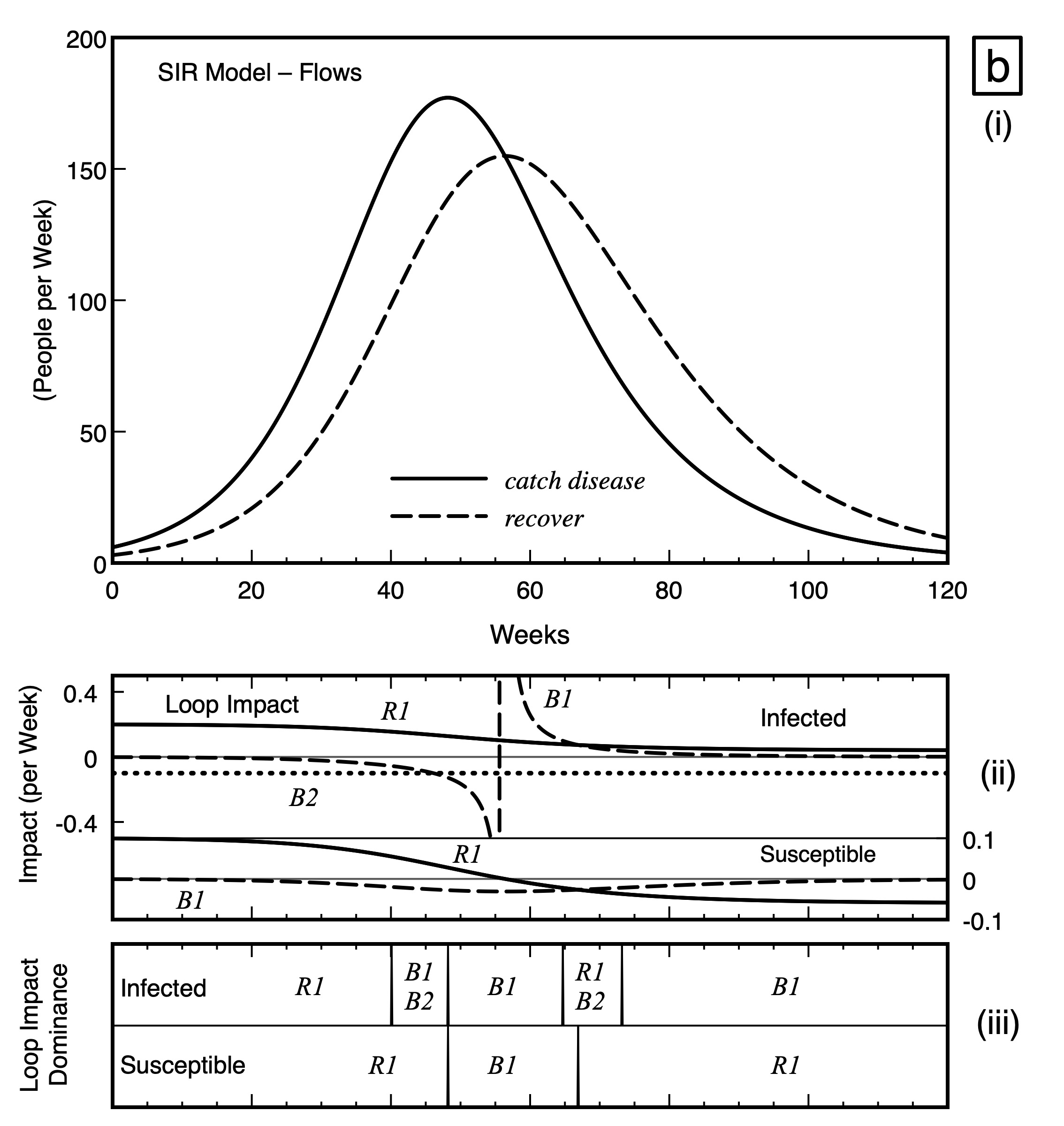}
      \end{minipage}
      \vspace{-25 pt}
       \caption{\small{Results of the SIR model.  (a) (i) Stocks; (ii) Loop impacts as a percentage of influence on the stock; (iii) Loop dominance transitions for loop impact.   (b). (i) Flows; (ii) loop impacts on \emph{Infected} and \emph{Susceptibles}; (iii) Loop dominance transitions for separate loop impact on the stocks. \emph{infection rate} = 0.2, \emph{recovery rate} = 0.1, \emph{Total Population} = 10,000,  \emph{initial Infected} = 30, \emph{initial Removed} = 0. }} \label{figure8.fig}
   \end{figure}

Our final example is the SIR model for the spread of an infectious disease, figure \ref{figure7.fig}. The system is conserved with a constant total population number. Thus,  the transmission rate \emph{catch disease} has only two loops \cite{sterman2000business,lyneis2007epidemic}, referred to as the \emph{mass action}  model \cite{hethcote1994thousand}. Each of the three loops in the model acts on two stocks. For example, $B_1$ acts as a first-order balancing loop on the susceptible stock and also acts exogenously on the infected stock via the inflow. This latter influence is an ``equal and opposite'' reaction to the former \cite{hayward2017newton}.  The average loop impact is the average of the magnitudes of these two, with the sign determined by the polarity of the loop gain (see appendix). By contrast, \citeasnoun{hayward2012model} pictured the influence of $B_1$ on the infected as part of a second-order loop formed with $R_1$. The impacts are identical in both approaches. The exogenous approach is more transparent for a loop dominance explanation and for computing average loop impacts. The results of the model are given in figure \ref{figure8.fig}.

The number of infected rises to a peak then falls to zero before all the susceptibles catch the disease, figure \ref{figure8.fig}a(i). According to the average loop impact, the early accelerating phase of the stocks is dominated by $R_1$, their changes of curvature are largely due to $B_1$, with $B_2$ dominating the slow down to equilibrium,  figures \ref{figure8.fig}a(ii-iii).  This compares well with the application of loop impact on single stocks, figures \ref{figure8.fig}b(ii-iii). The impact of the three loops on the infected has similar transitions to the average loop impact. Here $B_1$ controls the infected around its maximum acting in an exogenous way via the shared flow \emph{catch disease}.(Compare figures \ref{figure8.fig}b(iii) with  \ref{figure8.fig}a(i)) The loop transitions on the susceptibles are similar, except that its slow down is control by $R_1$ acting exogenously, which has a balancing effect of the susceptibles in this phase. The  S-shaped behaviour of the removed is determined by $B_2$ alone as its impact is not constant, reflecting the exogeneity of its influence (see appendix).    Thus average loop impact has captured the stock behaviours reasonably well.

\section*{Conclusion}
The concept of average loop impact was introduced as a single number alternative to the individual impacts on each stock in a loop.  Average loop impact is able to give a system-wide analysis for loop dominance by providing a single measure of the effect of a loop a one set of loop transitions for a model. Although this approach appears simpler than the Loop Impact method as originally presented by \citeasnoun{hayward2014model}, the disadvantage is that the connection between loop measure and behaviour is less clear. Loop Impact provides loop transition diagrams for each stock, where the periods of acceleration and deceleration are clearly linked to the measures of the loops. It is inevitable that to fully describe the effect of a feedback loop, there need to be as many measures as there are stocks in the loop. To decide which method to use, a modeller needs to be clear what behaviour they are seeking to explain. The greater the clarity of  connection between structure and the behaviour of individual model elements, the more loop transition diagrams are need and the more complex the explanation. This is the approach of loop Impact. If a simpler explanation is required, then average loop impact  may provide it but at the expense of a clear definition of model behaviour in terms of graphs over time.

\section*{Appendix}

Firstly, it is shown how the average loop impact in the SIR model is computed. The SIR model, figure \ref{figure7.fig}, reduces to the differential equation:
\begin{eqnarray}
\dot{S} &=& - f = -\beta SI \label{S.eq} \\
\dot{I} &=& f - g= \beta SI - \gamma I \label{I.eq} \\
\dot{R} &=& g = \gamma I \label{R.eq}
\end{eqnarray}
where $\beta$ is the \emph{infection rate} divided by the \emph{total population}; $\gamma$ is the \emph{recovery rate}, $f,g$ are the two flows,  and $S,I,R$ are the three stocks. 

Loop impacts are obtained from these equations using pathway differentiation \cite{hayward2017newton}. For example, the impact of $B_1$ on $S$ is derived from (\ref{S.eq}): 
\begin{equation}
\St_{\underline{SfS}}(B_1) = -\frac{\partial f }{\partial S} = -\beta I \label{b1S.eq}
\end{equation}
 Using (\ref{I.eq}), the impact of $B_1$ on $I$ is: 
 \begin{equation}
 \St_{\underline{SfI}}(B_1) = \frac{\partial f }{\partial S}\frac{\dot{S}}{\dot{I}} = -\frac{\beta^2SI}{\beta S - \gamma}\label{b1I.eq}
\end{equation}
where the fraction $\dot{S}/\dot{I}$ is needed as the source stock is different from the target stock in this exogenous influence \citeaffixed{hayward2012model}{c.f. sec. 5.2}. The average loop impact of $B_1$ is obtained from (\ref{b1S.eq}--\ref{b1I.eq}) as 
\[
\St_{average}(B_1)=-\tfrac{1}{2}\left( \left|\St_{\underline{SfS}}(B_1)\right| + \left|\St_{\underline{SfI}}(B_1)\right| \right)
\] 

Secondly, it is shown how the removed stock exhibits S-shaped growth even though it is only  influenced by one loop, $B_2$, whose gain is constant. Although its loop impact on $I$ is constant, $\St_{\underline{IgI}}(B_2)= -\gamma$, its exogenous impact on the removed is more complex $\St_{\underline{IgR}}(B_2)= \gamma \dot{I}/\dot{R} = \beta S-\gamma$ as the source and target stocks differ. The curvature of $R$ reflects the behaviour of  $S$. $R$ starts by accelerating $\beta S> \gamma$, then transitions to deceleration as $S$ falls in value and the impact becomes negative.  The result is the removed's S-shaped behaviour even though it is only influenced by one first-order loop, figure \ref{figure8.fig}a(i).

\end{document}